\documentclass{aa}
\usepackage{graphicx}
\usepackage{txfonts}
\begin{document}
   \title{Narrow-line \ion{H}{i} and cold structures in the ISM}
   \author{U. Haud}
   \institute{Tartu Observatory, 61\,602 T\~oravere, Tartumaa, Estonia\\
              \email{urmas@aai.ee} }
   \date{Received \today; accepted \today}
   \abstract
      {In the \ion{H}{i} line profiles in the Leiden-Argentina-Bonn
         (LAB) all-sky database, we have found a population of very cold
         \ion{H}{i} clouds. So far, the role of these clouds in the
         interstellar medium (ISM) has remained unclear.}
      {In this paper, we attempt to confirm the existence of the
         narrow-line \ion{H}{i} emission (NHIE) clouds by using the data
         from the Parkes Galactic all-sky survey (GASS) and try to find
         their place among other coldest constituents of the ISM.}
      {We repeat the search of NHIE with the GASS data and derive or
         compile some preliminary estimates for the distribution,
         temperatures, distances, linear sizes, column and number
         densities, masses, and the composition of NHIE clouds, and
         compare these data with corresponding estimates for \ion{H}{i}
         self-absorption (HISA) features, the Planck cold clumps (CC),
         and infrared dark clouds (IRDC).}
      {We demonstrate that from LAB and GASS we can separate comparable
         NHIE complexes, and the properties of the obtained NHIE clouds
         are very similar to those of HISA features, but both of these
         types of clouds are somewhat warmer and more extended and have
         lower densities than the cores in the Planck CC and IRDC.}
      {We conclude that NHIE may be the same type of clouds as HISA,
         but in different observing conditions, in the same way as the
         Planck CC and IRDC are most likely similar ISM structures in
         different observing conditions and probably in slightly
         different evolutionary stages. Both NHIE and HISA may be an
         intermediate phase between the diffuse cold neutral medium and
         star-forming molecular clumps represented by the Planck CC and
         IRDC.}


   \maketitle

   \section{Introduction}

      In a series of papers, we have described the Gaussian
      decomposition of 21-cm line surveys (Haud \cite{Hau00}) and the
      use of the obtained Gaussians for the detection of different
      observational and reductional problems (Haud \& Kalberla
      \cite{Hau06}), for the separation of thermal phases in the
      interstellar medium (ISM; Haud \& Kalberla \cite{Hau07}), and for
      the studies of intermediate- and high-velocity hydrogen clouds
      (Haud \cite{Hau08}). Observational data for the decomposition were
      taken from the Leiden-Argentina-Bonn (LAB) all-sky database of
      \ion{H}{i} 21-cm line profiles, which combines the
      Leiden/Dwingeloo Survey (LDS, Hartmann \& Burton \cite{Har97}) and
      a similar southern sky survey (Bajaja et al. \cite{Baj05})
      completed at the Instituto Argentino de Radioastronom\'{i}a. The
      LAB database with its improved stray-radiation correction is
      described in detail by Kalberla et al. (\cite{Kal05}).

      In the latest paper in this series (Haud \cite{Hau10}, hereafter
      Paper~I), we tested our new algorithm for the separation of the
      clouds of similar Gaussians from the general database of the
      Gaussian parameters. For technical reasons, we focused mainly on
      the clouds of the narrowest Gaussians, and this led us to
      surprising results. We found in the sky a $80\degr$ long filament
      of narrow-lined \ion{H}{i} emission (NHIE) clouds and modeled this
      filament as part of a ring-like structure (Figs.~3 and 4 of
      Paper~I). According to the obtained model, the center of the ring
      is located in the direction $l = 236\fdg2 \pm 0\fdg9, b = -13\fdg2
      \pm 0\fdg3$. We also obtained a very rough distance estimate of
      $126 \pm 82~\mathrm{pc}$ for the ring center, reported that the
      linear radius of the ring of $113~\mathrm{pc}$ follows from this
      distance, and found that at its nearest point to the Sun the ring
      clouds are at about $33~\mathrm{pc}$ from the Sun.

      Based on the LAB data, the ring clouds are mostly represented by
      the strong narrow Gaussians, which are clearly distinguishable
      from the broader lines of \ion{H}{i} in the same sky region. The
      mean $\mathrm{FWHM}$ of the \ion{H}{i} lines in the ring clouds is
      only about $1.8~\mathrm{km\,s}^{-1}$. This is more than two times
      less than the average $\mathrm{FWHM} = 3.9~\mathrm{km\,s}^{-1}$,
      corresponding to the ordinary cold neutral medium (CNM, Haud \&
      Kalberla \cite{Hau07}) in interstellar space. At the same time,
      the actual line widths of the ring clouds may be even less than
      the estimated average. The channel separation of the LAB profiles
      is no less than $1.03~\mathrm{km\,s}^{-1}$, and owing to the
      saturation effects, possible substructure and turbulent motions
      inside the clouds, the actual line shapes need not be exactly
      Gaussian. All this may increase our line-width estimates, and
      therefore we may be sure that these lines are unusually narrow for
      Galactic \ion{H}{i} 21-cm emission. However, for weaker components
      and when approaching the Galactic plane, the non-uniqueness of the
      Gaussian decomposition, the blending with other \ion{H}{i}
      features, and the presence of traces of radio-frequency
      interferences (RFI) and stronger noise peaks in observed profiles
      become increasingly complicating factors for the separation of the
      narrow line components. Therefore, in Paper~I we did not discuss
      in detail the possible continuation of the ring to $b<20\degr$,
      where, according to our model, the clouds are located at greater
      distances from the Sun, and the corresponding line components
      become weaker and more blended with the ordinary emission of the
      Galactic disk.

      As mentioned above, the ring clouds found in Paper~I were among
      the most reliably detected narrow-lined features in LAB, but all
      together, we found 1\,336 cloud candidates (including those at
      $|b| \le 20\degr$), each with estimated mean $T_\mathrm{b} \ge
      1.0~\mathrm{K}$, $\mathrm{FWHM} \le 3.0~\mathrm{km\,s}^{-1}$,
      detected in at least seven neighboring sky positions and having
      $|V_\mathrm{LSR}| \le 15~\mathrm{km\,s}^{-1}$. The problems
      mentioned above make us admit that a fraction of these cloud
      candidates may be spurious features. Moreover, we also pointed out
      that the estimates of the Gaussian widths were only the upper
      limits for the actual line widths, and we could not discuss the
      physical properties of these clouds in detail. Therefore, the
      nature of these clouds remained rather enigmatic. In this paper,
      we try to use additional observations, data found in the
      literature, and indirect evidence from our own results to shed
      light on the question. To do this, in the next section we discuss
      the reliability of detecting of the narrow line components, then
      summarize the existing knowledge of these clouds, and in
      subsequent sections we review the properties of other objects,
      more or less similar to our NHIE clouds. In the final sections, we
      try to draw a coherent picture of the sequence of objects,
      representing different stages in the conversion of the diffuse CNM
      to star-forming, dense molecular cores, and present the
      conclusions.

   \section{Reliability of detections}

      In the Introduction we mentioned that our list of narrow-lined
      cloud candidates may be contaminated by the spurious features. The
      main sources of such contamination are
      \begin{itemize}
         \item the uncertainties in the Gaussian decomposition,
         \item blending of the narrow components with wider line
            emission,
         \item the presence of traces of radio-frequency interferences
            in the profiles, and
         \item the noise peaks in the profiles, which may be rather
            similar to weak narrow line emission.
      \end{itemize}
      Now we discuss these sources of contamination in greater detail,
      starting from the end of the list above.

      \begin{figure}
         \resizebox{\hsize}{!}{\includegraphics{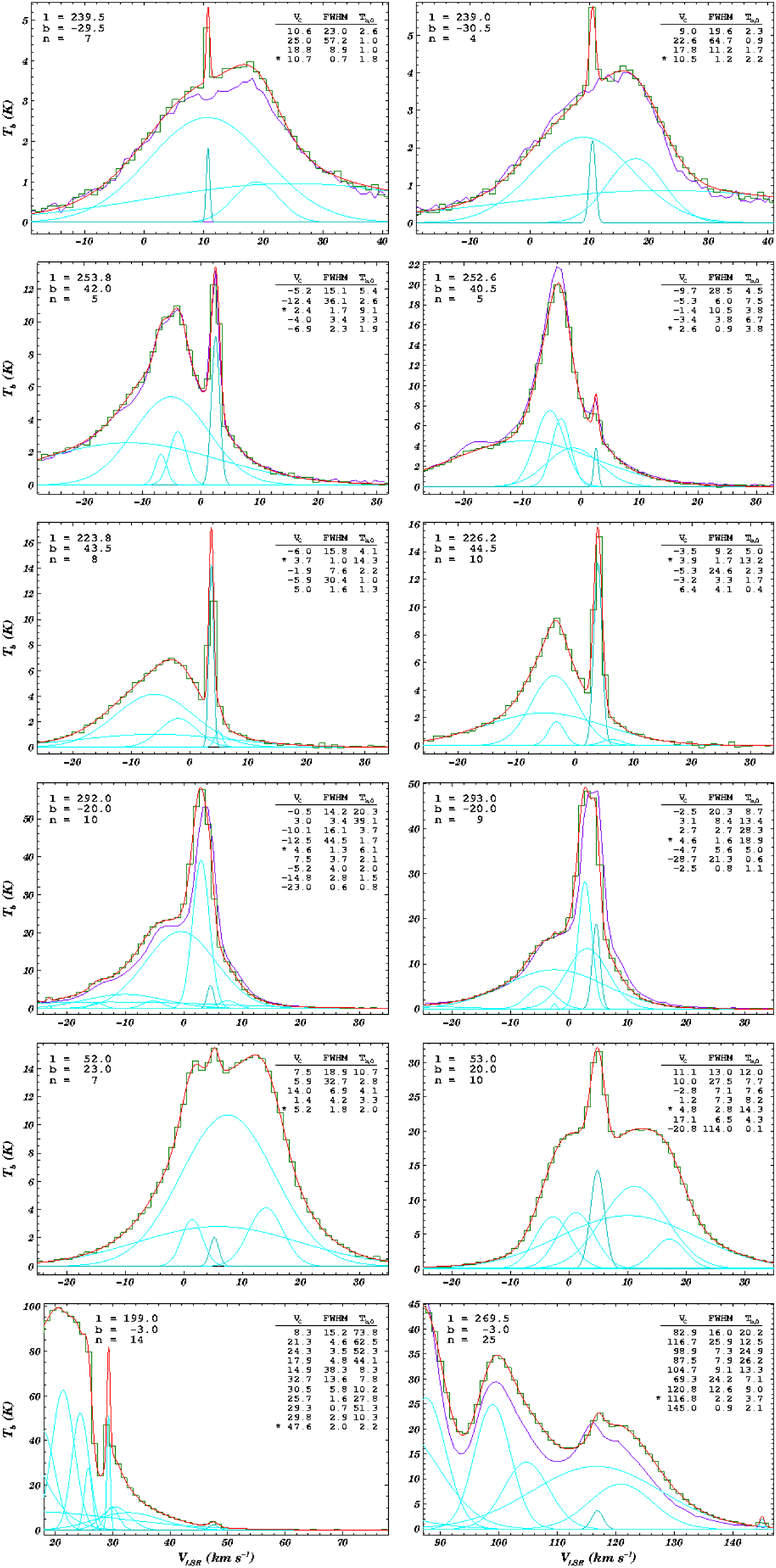}}
         \caption{Examples of profile fragments with narrow components.
            The observed LAB profiles are plotted by green stepped
            lines, individual Gaussian components by turquoise thin
            smooth lines, and the Gaussian representations of the
            profiles by red thick smooth lines. The dark violet
            piecewise linear lines represent the GASS observations. The
            numerical data in the upper left corner of each panel give
            the Galactic coordinates and the total number of Gaussians
            for this profile. In the upper right corner are the
            parameters of Gaussians, contributing to the model profile
            in the velocity range of the figure. The Gaussian component,
            belonging to the cloud under discussion, is drawn by a
            thicker dark turquoise line near the center of the velocity
            axis of each panel and marked with an asterisk in the table
            of Gaussian parameters. Row 1: the cloud candidate with the
            narrowest average line width. Row 2: the ring cloud with the
            narrowest average line width. Row 3: Verschuur's cloud A.
            Row 4: heavily blended components. Row 5: profiles from the
            comet-like structure, discussed in Sec. 3. Row 6: two clouds
            at higher velocities near the Galactic plane, used for
            Fig.~\ref{Fig04}.}
         \label{Fig01}
      \end{figure}
      As described in Haud (\cite{Hau00}), the Gaussian decomposition of
      the LAB profiles was adjusted so that on average once per two
      profiles some stronger noise peak was fitted with a Gaussian,
      which was then also included in the final database of the
      Gaussians. This approach improved the detection of weak signal
      features, but at the same time, in this way slightly more than
      $6\%$ of all Gaussians in our final database actually represent
      noise, and most of these components are rather narrow (in
      Fig.~\ref{Fig03}, discussed in greater detail later in this paper,
      they are concentrated around $(T_\mathrm{b}, \mathrm{FWHM}) =
      (0.5~\mathrm{K}, 0.9~\mathrm{km\,s}^{-1})$). Therefore, at first
      glance, it may seem that a considerable part of the cloud
      candidates may represent such noise features, but actually this is
      not the case. First of all, we have required from all cloud
      candidates to have a mean of $T_\mathrm{b} \ge 1.0~\mathrm{K}$,
      and this excludes most of the noise peaks from consideration.
      Moreover, the noise is random in its nature, and therefore, it is
      highly improbable that many neighboring profiles exhibit strong
      noise features at similar velocities. Since our cloud candidates
      all have similar components from at least seven neighboring sky
      positions, we believe that the contamination by the noise is
      actually negligible and may become somewhat considerable only in
      the case of Fig.~\ref{Fig04}, for which smaller clouds, detected
      at least in three neighboring sky positions, were also used.

\begin{figure}
         \resizebox{\hsize}{!}{\includegraphics{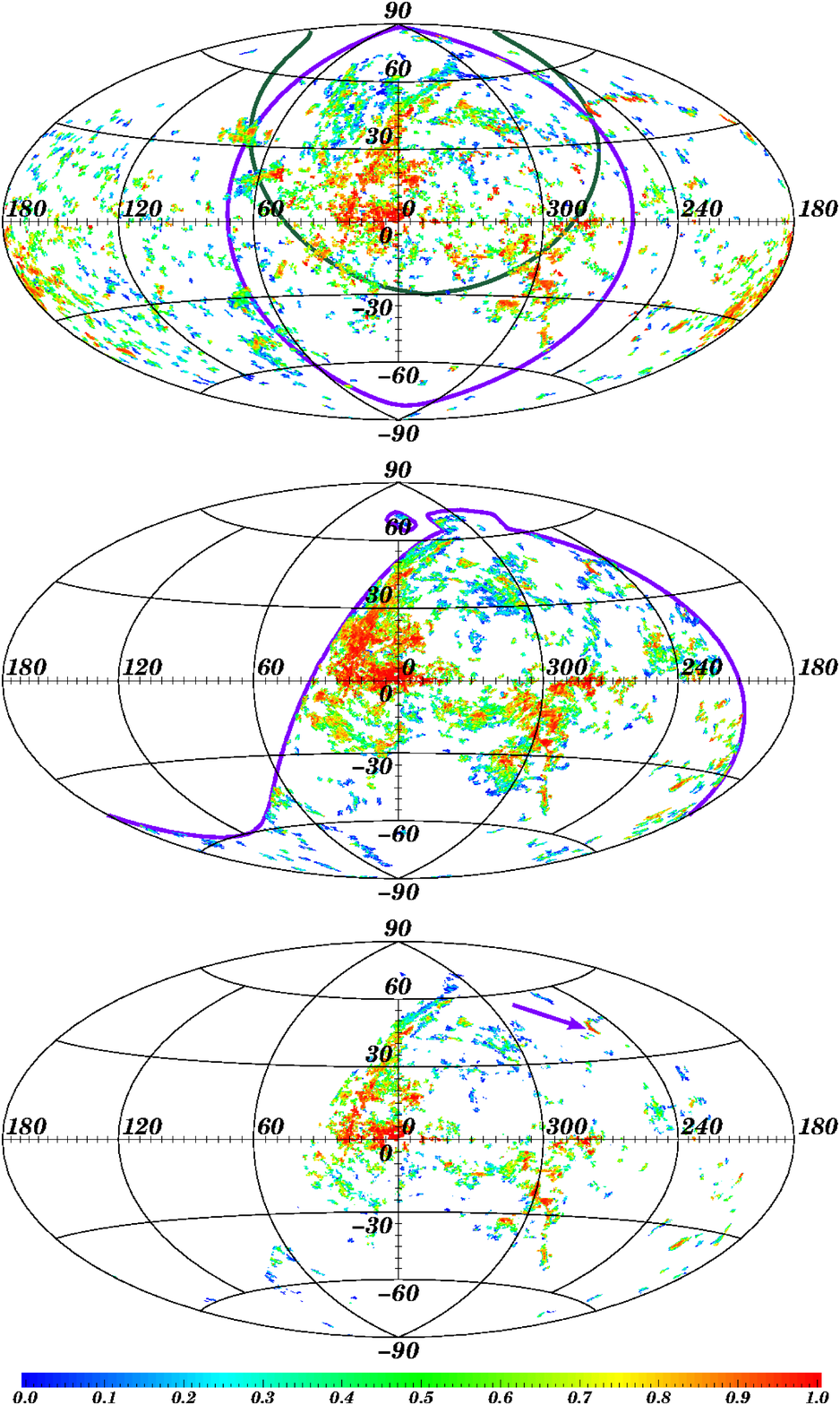}}
         \caption{The NHIE clouds from the LAB (upper panel), from the
            GASS (middle panel), and from both (lower panel). Every
            Gaussian, belonging to an NHIE cloud candidate, is
            represented with the color, corresponding to the sequence
            number of this component in the ascending list of the values
            of $T_\mathrm{b,0} / \mathrm{FWHM}^2$ of these Gaussians.
            The red color corresponds to the highest values of
            $T_\mathrm{b,0} / \mathrm{FWHM}^2$. The numbers on the color
            scale below the figure denote the fractions of the length of
            the list. The thick lines in the upper panel indicate the
            location of the expanding superbubble shells S1 (dark
            violet) and S2 (sea green). In the middle panel, the border
            of the region covered by the GASS is indicated. The arrow in
            the lower panel points to the cloud with the narrowest
            lines, reliably detected in both the LAB and GASS.}
         \label{Fig02}
      \end{figure}

      The situation is more complicated with RFI, since they are often
      considerably stronger than the noise and persist during certain
      time periods during the observations. In the case of the LAB, the
      sky was observed by $5\degr \times 5\degr$ fields. If the
      interference existed during a considerable part of the
      observations of one field and remained undetected in the process
      of the following data reduction, we may obtain nice ``clouds'' of
      RFI. These doubts are deepened by Fig. 3 of Paper~I, or the upper
      panel of Fig.~\ref{Fig02} here. Especially in the southern sky, we
      can see some ``clouds'' of approximately rectangular shape and
      size close to $5\degr \times 5\degr$ (e.g., the ``clouds'' near
      $(l, b) = (336\degr, -70\degr)$ and $(318\degr, -62\degr)$, but
      also others). As described in Paper~I, we applied special
      selection criteria to suppress such features, but it seems that
      these criteria have not been successful in all cases.

      This confirms that not all our candidates correspond to real
      clouds, but from the LAB alone, it is hard to distinguish which
      may be real and which may be spurious. However, the RFI is often
      specific to a particular observation time, location, or equipment.
      Therefore, the best check of the situation may be obtained by
      observing the same sky regions at different times and locations
      with different equipment. Such a possibility for the southern sky
      is provided by the Parkes Galactic all-sky survey (GASS;
      McClure-Griffiths et al. \cite{McC09}, the data of the second data
      release by Kalberla et al. \cite{Kal10} are available at
      http://www.astro.uni-bonn.de/hisurvey), and this check confirms
      that not only the rectangular ``clouds'' referred to above, but
      also some more naturally looking ones are most likely caused by
      RFI.

      The example of the situation is given in the first row of
      Fig.~\ref{Fig01}. From the LAB data we found at $(l, b) =
      (239\degr, -31\degr)$ a ``cloud'' with narrow line width. The line
      profiles from this region exhibit a clear narrow feature at about
      $V_\mathrm{LSR} = 10.7~\mathrm{km\,s}^{-1}$. Similar features are
      detected in nine neighboring profiles, and Fig.~\ref{Fig01}
      illustrates two of them together with corresponding Gaussian
      decompositions. However, nothing similar was found in GASS. As a
      result, we must recognize that this ``cloud'' is most likely a
      spurious feature. At the same time, as we can see from the second
      row of Fig.~\ref{Fig01}, the reality of one ring cloud is
      confirmed well by the GASS data and, as discussed later, some more
      of them, located outside the region which is covered by the GASS,
      are confirmed by other independent observations (e.g., the one in
      the third row of Fig.~\ref{Fig01}).

      Many narrow Gaussians are heavily blended by the wider ones, and
      in such cases, even the direct comparison of the observed profiles
      from the LAB and GASS may not give a conclusive result (the fourth
      row of Fig.~\ref{Fig01}). Such blended components may easily be
      the artifacts of the Gaussian decomposition, since it is well
      known that often the Gaussian decomposition is not unique, and
      several quite different solutions may approximate the observed
      profiles almost equally well. The decomposition provides no
      satisfactory means of choosing between these solutions, while
      other equally good or even better ones may not be found at all.
      Such nonuniqueness affects the heavily blended components most
      seriously, as in these cases even small changes in the data or in
      the decomposition process may lead to completely different
      decomposition results. Therefore, to check how prone our results
      are to such uncertainties, the most conclusive solution is to
      repeat the search of the clouds using more or less independent
      observations and a more or less different decomposition
      algorithm.

      As mentioned above, in the southern sky the possibility of using
      independent data is offered by the GASS. For the LAB and GASS,
      different observing instruments (30-m dish of the Instituto
      Argentino de Radioastronom\'{i}a and the Parkes 64 m radio
      telescope, respectively) and techniques (pointed observations for
      the LAB against on-the-fly observations in the GASS) were used.
      Compared to the LAB, the GASS has considerably better spatial
      resolution and slightly better velocity resolution and
      sensitivity. For the northern sky, the observations, similar to
      the GASS, are going on in Germany (the Effelsberg-Bonn \ion{H}{i}
      Survey or EBHIS; Kerp et al. \cite{Ker11}), but the corresponding
      results are not yet available. Therefore, it seems interesting to
      decompose the GASS data and to compare the results with the
      decomposition of the LAB. For the northern sky we have to hope
      that the outcome of any future comparison of the EBHIS and LAB
      will in general be similar.

      For decomposition, the GASS data were prepared in the HEALPix grid
      (G\'{o}rski et al. \cite{Gor05}) with $N_\mathrm{side} = 1024$ by
      P. Kalberla, and the obtained 6\,655\,155 profiles were decomposed
      with the modified version of our decomposition program into
      60\,349\,584 Gaussians. We used the original version of the same
      program (Haud \cite{Hau00}) also for the decomposition of the LAB
      data, and therefore we cannot claim that the decomposition
      algorithms, which were applied to the LAB and GASS, were
      completely independent, but there were nevertheless important
      differences. First of all, the algorithms for finding the initial
      approximations for the decomposition of each profile were
      completely different. When processing the LAB, for the
      decomposition of a new profile the result of the earlier
      decomposition of one of its neighbors was used as the initial
      approximation. In the case of the GASS, the decomposition of each
      profile was started independently of all the others with only one
      Gaussian roughly fitting the main peak of the given profile.
      During the following decomposition, the weighting of the profile
      channels was also made differently for the LAB and GASS. In the
      LAB, we had just one profile for most sky positions, and therefore
      the dependence of the uncertainties of the channel values on the
      signal strength had to be estimated semitheoretically (discussed
      in detail in Haud \cite{Hau00}). In the GASS, typically 40
      individual spectra contributed to every final resolution element.
      Therefore, it was possible to estimate, for every final profile,
      the uncertainties of all channels from the mutual deviations of
      the contributing profiles and from the presence of the flagged
      channels (Kalberla et al. \cite{Kal10}). Besides these main
      differences in the decomposition algorithms, there were also some
      others (a new path for proceeding through the survey profiles, a
      new definition of the neighborhood of all profiles, etc.), but
      their influence on the final decomposition results is probably
      less severe.

      Owing the described differences in the decomposition processes of
      the LAB and GASS, it seems plausible that in ambiguous cases we
      will get rather different sets of Gaussians for the profiles of
      these two surveys, and similar Gaussians will indicate the real
      features, detected by both surveys. For the final comparison of
      the decomposition results, we decided also to repeat the
      cloud-finding procedure applied to the LAB Gaussians with the GASS
      data. We carried this procedure out in exactly the same way as for
      the LAB with only one exception. The HEALPix grid for
      $N_\mathrm{side} = 1024$ is much denser than the grid used for the
      observations of the LAB, and therefore, if we would like to search
      in both surveys for the clouds of approximately the same sizes,
      the requirement on the number of neighboring sky positions with
      similar narrow components must be considerably increased for the
      GASS. From the comparison of the total number of the profiles in
      the LAB with the number of resolution elements in HEALPix, we
      concluded that seven neighboring positions in the LAB correspond
      to 464 positions in the GASS and considered further only the GASS
      clouds of at least this size.

      The results of the comparison are presented in Fig.~\ref{Fig02}.
      The color of the points in this figure represents the values of
      the rather arbitrarily chosen parameter $T_\mathrm{b,0} /
      \mathrm{FWHM}^2$. The justification for this choice is that this
      parameter has the highest values for the strongest and narrowest
      components, which are the most interesting ones in the context of
      our discussion. However, as the actual values of the parameter are
      not very important, we rescaled them for every part of
      Fig.~\ref{Fig02} in the following way. First, we found which
      clouds satisfy all our selection criteria, then we calculated
      $T_\mathrm{b,0} / \mathrm{FWHM}^2$ for every Gaussian of these
      clouds, sorted the Gaussians in ascending order of the parameter
      values, and colored them according to their sequence number in the
      ordered list. In this way, in each panel the $10\%$ of the
      strongest and narrowest Gaussians are drawn with red color, next
      $10\%$ are orange, next $10\%$ yellow, etc. The upper panel of
      Fig.~\ref{Fig02} reproduces all clouds from the LAB, the middle
      panel is for the results from the GASS and the lowest panel
      represents the product of the upper two. This means that in the
      lowest panel only those clouds are represented that were found in
      both the LAB and GASS. We see that all three panels of
      Fig.~\ref{Fig02} are remarkably similar and for the southern sky
      we may find all main features in all these panels. This is not the
      result that we expect if most of our narrow Gaussians are
      artifacts of the Gaussian decomposition of ambiguous cases of
      heavily blended profile features. Therefore, we conclude that
      despite some spurious features in the list of the cloud
      candidates, most of the clouds, found by us in the LAB and then
      found again in the GASS, are real. However, if so, we may also ask
      about their physical nature.

   \section{Narrow-line neutral hydrogen emission}

      When we started to search for references to objects similar to
      those found in our study, the results were rather scarce. The
      situation was clarified by the recent statement of Gibson
      (\cite{Gib11}): ``Historically only a few NHIE features were known
      (Knapp \& Verschuur \cite{Kna72}; Goerigk et al. \cite{Goe83}),
      but increasingly sophisticated spectral decompositions have
      revealed more (Verschuur \& Schmelz \cite{Ver89}; P\"oppel et al.
      \cite{Pop94}; Haud \cite{Hau10})''. Of all these papers, Knapp \&
      Verschuur (\cite{Kna72}) studied a cloud, which is also a part of
      the filament discussed by us in Paper~I. Goerigk et al.
      (\cite{Goe83}) studied the Draco cloud, which we also found in our
      analysis, but because according to Goerigk et al. (\cite{Goe83}),
      its mean $\mathrm{FWHM} > 4.0~\mathrm{km\,s}^{-1}$ (the velocity
      resolution of their data was $1.69~\mathrm{km\,s}^{-1}$) and
      $V_\mathrm{LSR} \approx -22~\mathrm{km\,s}^{-1}$, it is not among
      the 1\,336 cloud candidates discussed in Paper~I.

      Verschuur \& Schmelz (\cite{Ver89}) presented the results of
      observations of \ion{H}{i} emission profiles in 180 directions
      over the northern sky and noticed that the peak in the histogram
      of the line-width distribution occurs at $3~\mathrm{km\,s}^{-1}$.
      Their observations were made with the channel bandwidth of
      $0.258~\mathrm{km\,s}^{-1}$. Therefore, this NHIE most likely
      corresponds to normal CNM (mean $\mathrm{FWHM} =3.9 \pm
      0.6~\mathrm{km\,s}^{-1}$ according to Haud \& Kalberla
      \cite{Hau07}). We can also offer a similar comment on the results
      by P\"oppel et al. (\cite{Pop94}), who made a systematic
      separation of the CNM from the warm neutral medium (WNM). Using
      the data with a velocity resolution of about
      $2~\mathrm{km\,s}^{-1}$, and considering only the profile peaks
      with $\mathrm{FWHM} \ge 6~\mathrm{km\,s}^{-1}$, they found the
      mean $\mathrm{FWHM} \approx 10.6~\mathrm{km\,s}^{-1}$ for the CNM
      and about $23.5~\mathrm{km\,s}^{-1}$ for the WNM. Therefore, their
      NHIE most likely corresponds already to our line-width group 2 in
      the thermally unstable regime (mean $\mathrm{FWHM} =11.8 \pm
      0.5~\mathrm{km\,s}^{-1}$; Haud \& Kalberla \cite{Hau07}).

\begin{figure}
         \resizebox{\hsize}{!}{\includegraphics{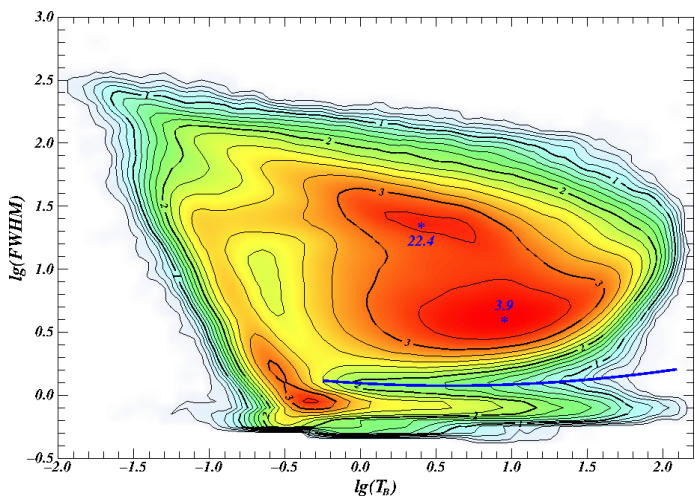}}
         \caption{Frequency distribution of the parameter values in the
            $(\lg(T_\mathrm{b,0}), \lg(\mathrm{FWHM}))$ plane for all
            Gaussians obtained in the decomposition of the LAB.
            Isodensity lines are drawn on the scale of $\lg(N+1)$ with
            the interval of 0.2. The two main maxima, corresponding to
            WNM and CNM, are labeled with the values of corresponding
            Gaussian $\mathrm{FWHM}$. The blue thick solid line
            represents the selection criterion defined by Eq. 4 of Haud
            \& Kalberla (\cite{Hau06}).}
         \label{Fig03}
      \end{figure}

      All this indicates that the notion of NHIE has so far been used
      almost for any \ion{H}{i}, colder than the WNM, and our clouds are
      among the most narrow-lined ones of the NHIE objects discussed
      above. From the frequency distribution of the Gaussian widths and
      heights, it may even seem that these very narrow-lined \ion{H}{i}
      emission (VNHIE?) clouds may belong to the population distinct
      from the CNM (Fig.~\ref{Fig03}, below the thick blue line).
      However, this conclusion is most likely an artifact of the
      observations and data reduction. We modeled the observations of
      Gaussian-shaped lines of different widths and realistic noise with
      a velocity channel separation typical of the LDS. After Gaussian
      decomposition of these ``observations'', we found that all lines
      are statistically well reproduced by such a procedure down to the
      line width of about $\mathrm{FWHM} \approx
      0.77~\mathrm{km\,s}^{-1}$. Below this limit the obtained widths of
      Gaussians become independent of the original line widths, and most
      decomposition results concentrate around the value of
      $0.77~\mathrm{km\,s}^{-1}$, with the sharp boundary of the
      distribution at $0.54~\mathrm{km\,s}^{-1}$. The same behavior is
      also observed in Fig.~\ref{Fig03}. Moreover, the frequency
      concentration below the thick line in Fig.~\ref{Fig03} is also
      enhanced by the contamination of the survey by the radio
      interferences, and most of the weakest Gaussians at these widths
      are due to the highest noise peaks in the data, which were
      considered by the decomposition program as a possible signal (Haud
      \& Kalberla \cite{Hau06}). For these reasons, we continue to use
      the acronym NHIE for our clouds, but at least in this paper we
      mean the VNHIE by it.

      Probably the most thoroughly studied object among our clouds is
      the one discovered by Verschuur (\cite{Ver69}) as cloud A at $(l,
      b) \approx (226\degr, 44\degr)$, and studied in more detail later
      by Verschuur \& Knapp (\cite{Ver71}), Knapp \& Verschuur
      (\cite{Kna72}), Crovisier \& Kaz\`es (\cite{Cro80}), Crovisier et
      al. (\cite{Cro85}), Heiles \& Troland (\cite{Hei3b}), Meyer et al.
      (\cite{Mey06}), and Peek et al. (\cite{Pe11b}). Already in his
      first paper on cloud A, Verschuur (\cite{Ver69}) stated that the
      cloud must have a kinetic temperature $T_\mathrm{k} \le
      30~\mathrm{K}$. Later authors have agreed with this estimate, and
      recent papers explain the observed line width as a result of the
      kinetic temperature $20_{-8}^{+6}~\mathrm{K}$, and the
      one-dimensional RMS turbulent velocity of $0.37 \pm
      0.08~\mathrm{km\,s}^{-1}$ (Meyer et al. \cite{Mey06}). Up to now,
      this cloud A has contained the narrowest \ion{H}{i} emission lines
      ever discovered, and it is also part of our ringlike ribbon of
      clouds (example profiles in the third row of Fig.~\ref{Fig01}),
      but it may not be the cloud with the narrowest lines in this
      ribbon. According to our estimates, the cloud (example profiles in
      the second row of Fig.~\ref{Fig01}) at the highest Galactic
      longitude tip of the stream may have even narrower lines. However,
      even now Verschuur's cloud A presents the narrow components, which
      are among the strongest and most securely identifiable of all
      similar components discussed in this paper.

      For other cloud candidates in our sample, we do not have such
      temperature estimates, and we even do not have enough information
      for obtaining them. In the Introduction we pointed out why our
      line-width estimates may be only the upper limits to the actual
      line widths. To obtain the temperatures from line widths, we must
      additionally know how much of the real line width is caused by the
      temperature and what is contributed by turbulence. We do not have
      this information. However, Heiles \& Troland (\cite{Hei3b}) have
      established a correlation between the upper limit on the kinetic
      temperature, $T_\mathrm{k,max}$, as estimated from the fitted
      width of the Gaussian component and the spin temperature
      $T_\mathrm{s}$ (their Eq. 15a). From our decomposition of the LAB
      data we found the average line width of the Gaussians belonging to
      cloud A, of about $1.6~\mathrm{km\,s}^{-1}$. Using this result in
      the correlation and taking into account that
      $T_\mathrm{k,max}=21.86 \mathrm{FWHM}^2$ and that for CNM the spin
      temperature is equal to the kinetic temperature (Heiles \& Troland
      \cite{Hei3a,Hei3b}), we obtain for this cloud $25~\mathrm{K}$.
      This is already within the errors of the more accurate estimate
      above.

      At the same time, the correlation by Heiles \& Troland
      (\cite{Hei3b}) was established for $T_\mathrm{k,max}$, estimated
      from the Gaussian fit of the opacity profiles, but we have used
      the decomposition of the emission lines. It is known from the
      equation of transfer that the emission line of an isolated,
      single, homogeneous \ion{H}{i} cloud may be represented well by a
      Gaussian only when its optical depth $\tau \ll 1$. For cloud A,
      Verschuur \& Knapp (\cite{Ver71}) already found that the shapes of
      the narrow lines were not a Gaussian, and they managed to fit
      these shapes with the values of the maximum optical depth between
      1 and 2. We fitted these non-Gaussian lines with a single
      Gaussian, and if we disregard that the criteria of isolation and
      homogeneity are essentially never met, we may calculate that for
      the optical depths cited above, such a Gaussian fitting results in
      the $13-25\%$ overestimation of the width of corresponding opacity
      profiles. Taking this into account, we reach the temperature
      estimates in the range $18 < T_\mathrm{k} < 21~\mathrm{K}$, in
      excellent agreement with the results of Meyer et al.
      (\cite{Mey06}). Therefore, we expect that it is possible in this
      way to obtain at least crude estimates of the temperatures of all
      clouds in our sample.

      The most interesting are the coldest clouds. The cloud candidate
      with the narrowest mean line width, $\mathrm{FWHM} \approx
      1.18~\mathrm{km\,s}^{-1}$, in our all-sky sample is located at
      $(l, b) \approx (239\degr, -31\degr)$. However, as mentioned
      above, when discussing the first row of Fig.~\ref{Fig01}, this
      cloud has not been confirmed by the GASS data. The cloud with the
      narrowest lines, strong signal, and clear confirmation from the
      GASS is at $(l, b) \approx 254\degr, 42\degr$ (the second row of
      Fig.~\ref{Fig01}, arrow in Fig.~\ref{Fig02}) and has the mean line
      width, $\mathrm{FWHM} \approx 1.25~\mathrm{km\,s}^{-1}$. From the
      correlation, established by Heiles \& Troland (\cite{Hei3b}), this
      line width corresponds to the temperature $T_\mathrm{k} =
      18~\mathrm{K}$. As described above, this estimate is most likely
      higher than the actual kinetic temperature of the cloud, which may
      be somewhere around $12~\mathrm{K}$ for the $25\%$ line-width
      correction.

      The upper limit of the temperatures of the clouds in our sample is
      not of special interest, since the limiting mean line width of the
      clouds in the sample was chosen rather arbitrarily. Nevertheless,
      the condition $\mathrm{FWHM} \le 3.0~\mathrm{km\,s}^{-1}$
      corresponds to $T_\mathrm{k} \le 66~\mathrm{K}$ with the same
      comments as for the lower temperature limit. Also applying the
      $25\%$ line-width correction here, this temperature becomes equal
      to $47~\mathrm{K}$. These rough estimates from the whole-sky data
      are in good agreement with the more accurate results by Dickey et
      al. (\cite{Dic03}) for a small test field. They find that clouds
      with temperatures below $40~\mathrm{K}$ are common (though not as
      common as warmer clouds with $40-100~\mathrm{K}$), with a long
      tail of the distribution reaching down to temperatures below
      $20~\mathrm{K}$.

      For the ring clouds, we also have some distance information. From
      our model in Paper~I, it follows that the clouds of the observed
      part of the ring are at distances $33-63~\mathrm{pc}$ from the
      Sun. At these distances their linear diameters, corresponding to
      the largest angular separation of the observed cloud points, are
      in the range $0.9-6.9~\mathrm{pc}$. The total length of the ribbon
      of the clouds is about $66~\mathrm{pc}$. At the same time,
      according to our rather rough estimates, Verschuur's cloud A is at
      the distance of about $34~\mathrm{pc}$ from the Sun. More
      recently, Peek et al. (\cite{Pe11b}) have found that this cloud is
      most likely in the distance interval of about
      $11.3-24.3~\mathrm{pc}$. Proceeding from this distance estimate,
      we may conclude that the cloud sizes in our ribbon are about
      $0.3-4.1~\mathrm{pc}$, and the total length of the discussed cloud
      complex is $21.9-47.0~\mathrm{pc}$.

      \begin{figure}
         \resizebox{\hsize}{!}{\includegraphics{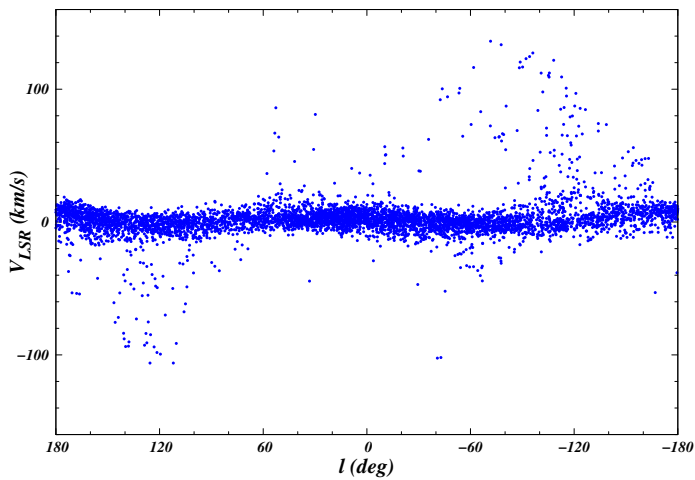}}
         \caption{The $l-V$ diagram for NHIE clouds (blue dots),
            observed in at least three neighboring sky positions.}
         \label{Fig04}
      \end{figure}
      Information on the distances of other clouds in our whole sky
      sample, identified in Paper~I, is practically missing. However,
      these objects have small line widths and, consequently, low
      temperatures and turbulent motions. If they have large linear
      dimensions, they cannot exist in such a state for a long time, and
      most likely they are relatively small clouds. At the same time,
      some of them cover up to 150 square degrees in the sky. This is
      only possible if they are relatively nearby.

      This also agrees with the fact that all clouds of this sample have
      the line-of-sight velocities $|V_\mathrm{LSR}| \le
      15~\mathrm{km\,s}^{-1}$. At the distances beyond the local spiral
      arm, the differential rotation of the Galactic disk would produce
      higher velocities at least in some regions near the Galactic
      plane. However, in Paper~I we also mentioned that in the initial
      selection 44 clouds with higher velocities were detected near the
      Galactic plane. These clouds were in the regions of the sky, where
      their velocities are naturally explained by the Galactic
      differential rotation. If we relax our selection criteria on the
      cloud sizes and velocities, we will get more clouds with higher
      velocities and their velocity distribution starts to resemble the
      usual $l-V$ diagram for the Galactic gas (Fig.~\ref{Fig04}).
      Unfortunately, such relaxation also increases the probability of
      confusion of the real NHIE clouds with the ``clouds'' of
      Gaussians, representing the traces of the removed radio
      interferences or observational noise. Nevertheless, we believe
      that we may conclude that the narrow lined clouds are detectable
      in the LAB at distances from some or some tens of parsecs up to
      several kiloparsecs. Two examples of the line profiles near the
      Galactic plane, containing the narrow components at higher
      velocities are given in the last row of Fig.~\ref{Fig01}.

      Something could also be said about the locations and shapes of
      NHIE clouds on the basis of their sky maps (Figs.~3 and 5 of
      Paper~I or the upper panel of Fig.~\ref{Fig02} here). Besides the
      long filament of the ring clouds discussed above, the most
      prominent clouds or complexes seem to be four structures at
      approximately $l = 70\degr$ and $b = -50\degr$, $-30\degr$,
      $15\degr$, $35\degr$, and the wide bands of clouds from $(l, b) =
      (20\degr, 0\degr)$ to $(290\degr, 70\degr)$ and from $(l, b) =
      (320\degr, -10\degr)$ to $(260\degr, -55\degr)$. All these
      apparently largest structures seem to be somehow related to
      expanding spherical superbubble shells S1 and S2, which are part
      of Loop~I superbubble. These shells were introduced by Wolleben
      (\cite{Wol07}) to explain the results of the Dominion Radio
      Astrophysical Observatory Low-Resolution Polarization Survey
      (Wolleben et al. \cite{Wol06}) in the North Polar Spur (NPS)
      region.

      The ring structure, modeled in Paper~I, nearly perpendicularly
      intersects shell S2 in the direction $l = 256\degr, b = 43\degr$
      and extends outward of this shell towards the lower Galactic
      longitudes. The gas stream from $(l, b) = (320\degr, -10\degr)$ to
      $(260\degr, -55\degr)$ is similar to the ring clouds. The stream
      projects onto the shell S1, and starts near the outer boundary of
      the shell S2, extending nearly perpendicular away from this shell.
      Both the ring clouds and the stream have their lowest
      line-of-sight velocities near S2, and the velocities increase when
      moving away from the shell. At the same time, the southern stream
      is much more irregular than the ring clouds in the northern sky.
      Nevertheless, both these structures may be parts of some shell
      structures.

      Most of these other NHIE structures touch or intersect the visible
      boundaries of these shells as well, and the structure nearest to
      $l = 0\degr$ resembles in its location, shape, and size the NPS,
      the brightest filament of Loop~I, which is a large circular
      feature in the radio continuum sky (Quigley \& Haslam
      \cite{Qui65}; Fig.~5 of Paper~I or Fig.~\ref{Fig02} here). As the
      distance of the NPS from the Sun is estimated to be about
      $120~\mathrm{pc}$ (Bingham \cite{Bin67}, Spoelstra \cite{Spo72}),
      the similarity of this stream to NPS may also serve as an
      additional hint that the NHIE clouds are probably rather nearby
      features. Many authors (e.g. Berkhuijsen et al. \cite{Ber71},
      Salter \cite{Sal83}, Egger \cite{Egg95}, Breitschwerdt \& Avillez
      \cite{Bre06}) have argued that the radio loops are correlated with
      expanding gas and dust shells, energized by supernovas or stellar
      winds. They may be interesting objects where a hot gas interacts
      with a cool and dense \ion{H}{i} interstellar medium (Park et al.
      \cite{Par07}).

      Many of the NHIE features discussed here have elongated or
      filamentary shapes, and some of them also have an interesting
      internal structure. The most remarkable is the cloud complex
      around $(l, b) = (60\degr, 15\degr)$ (Fig.~6 of Paper~I). Its
      densest parts have the highest velocities and resemble the stream
      of gas, nearly perpendicular to the borders of the images of the
      S1 and S2 shells. This relatively dense and fast-moving gas is
      mostly surrounded by an envelope with lower observed line-of-sight
      velocities and surface densities. The most narrow-lined (the
      coolest?) gas is located in front of the head of this comet-like
      stream (in the region, closest to the S1 and S2). In central
      regions of the cloud, the narrow features are much stronger and
      slightly wider. The examples of the profile segments of the
      coolest gas and the gas near the center of the head of the
      comet-like structure are given in the fifth row of
      Fig.~\ref{Fig01}. The traces of a similar cometary structure may
      also be observed in the cloud around $(l, b) = (70\degr,
      -50\degr)$.

      Finally, if we know the \ion{H}{i} 21-cm line profiles, it is
      possible to calculate neutral hydrogen column densities for the
      optically thin clouds. Knowing the sizes of the clouds, it is
      possible to convert these column densities to spatial densities
      and the \ion{H}{i} mass estimates of the clouds. Using the
      temperature estimates, we can determine the pressure and so on.
      However, we have already mentioned that at least some line
      profiles are seriously affected by the limited velocity resolution
      of the LAB data, by blending the narrow features with other
      \ion{H}{i} gas, and by problems in the Gaussian decomposition.
      Moreover, according to Verschuur \& Knapp (\cite{Ver71}), the
      clouds are not optically thin. Later, Peek et al. (\cite{Pe11b})
      found that their double unsaturated Gaussian model is the best for
      narrow lines of cloud A. In any case, since we have fitted these
      lines with single Gaussians, the parameters of these Gaussians
      cannot be representative of the actual physical properties of the
      clouds. Moreover, our temperature estimates are only based on the
      statistical correlation between absorption line widths and kinetic
      temperatures which is applied to emission line widths. The
      distances follow from the correlation between the line widths and
      linear dimensions of the molecular clouds which is applied to
      \ion{H}{i} clouds. Therefore, we must conclude that the obtained
      estimates are only very rough ones, and their combinations are
      even more questionable.

      However, such estimates based on better data were made for
      Verschuur's cloud A by Heiles \& Troland (\cite{Hei3b}), Meyer et
      al. (\cite{Mey06}), and Peek et al. (\cite{Pe11b}). They find for
      this one cloud the \ion{H}{i} column densities ranging up to $2.5
      \times 10^{20}~\mathrm{cm}^{-2}$, the density of
      $150-320~\mathrm{cm}^{-3}$, and the mass $0.235-1.07~M_{\sun}$.
      This cloud has been also observed in 18-cm OH and 2.6-mm CO lines,
      but without positive detections (Crovisier \& Kaz\`es
      \cite{Cro80}). From the $100~\mu\mathrm{m}$ infrared observations
      Peek et al. (\cite{Pe11b}) conclude that cloud A has either a
      lower-than-expected overall dust grain density or a population
      that has somewhat larger grains than is typical for the ISM.

   \section{Neutral hydrogen self-absorption}

      More commonly than in emission, the cold \ion{H}{i} is identified
      in absorption against either a continuum source (\ion{H}{i}
      continuum absorption = HICA) or other line emission (\ion{H}{i}
      self-absorption = HISA). HICA is the method of choice for
      exploring CNM properties, because it has fewer radiative transfer
      unknowns (Heiles \& Troland \cite{Hei3b}). However, HICA sight
      lines are discrete and well separated in present surveys, so that
      individual clouds are often sampled only once, or missed entirely.
      Therefore, HISA is the preferred method for mapping detailed CNM
      structure in absorption. Moreover, HISA backgrounds are typically
      not as bright as HICA backgrounds, and so HISA observations are
      more focused on the coldest \ion{H}{i} (Gibson \cite{Gib11}).

      Neutral hydrogen self-absorption was discovered (Heeschen
      \cite{Hee54,Hee55}) only three years after the first detection of
      \ion{H}{i} emission from interstellar hydrogen in the Galactic
      plane (Ewen \& Purcell \cite{Ewe51}). HISA was observed in
      \ion{H}{i} profiles as a very narrow absorption dip of only
      $1-4~\mathrm{km\,s}^{-1}$ wide (Knapp \cite{Kna74}). Much of the
      earlier work used single-dish observations and achieved the
      angular resolutions of up to $3 \farcm 2$ with the 1\,000-foot
      Arecibo telescope (Baker \& Burton \cite{Bak79}; Bania \& Lockman
      \cite{Ban84}). More recently, such interferometric surveys as the
      Canadian Galactic Plane Survey (CGPS; Taylor et al. \cite{Tay03}),
      Southern Galactic Plane Survey (SGPS; McClure-Griffiths et al.
      \cite{McC05}), and VLA Galactic Plane Survey (VGPS; Stil et al.
      \cite{Sti06}) have reached the angular resolutions of $1 \arcmin$
      and a velocity sampling around $0.82~\mathrm{km\,s}^{-1}$. These
      surveys have provided a wealth of data in which to look for and
      study the HISA phenomenon (Kavars et al. \cite{Kav05}).

      From these data, the physical parameters of some HISA features
      have been determined. The results have been reviewed by Kavars et
      al. (\cite{Kav05}) and Gibson (\cite{Gib11}), among others. From
      these reviews and other sources, it follows that HISA features are
      observable around the Sun in the distance range from about
      $0.1~\mathrm{kpc}$ up to at least $7~\mathrm{kpc}$, that their
      linear dimensions are about $0.029-3.6~\mathrm{pc}$, and that the
      dimensions of the cloud complexes extend up to
      $10-110~\mathrm{pc}$ (Table~2 from Kavars et al. \cite{Kav05},
      excluding the clouds with only the upper limits for distances, and
      the Local Filament from Table~1 of Gibson et al. \cite{Gib00}).
      The estimated spin temperatures of HISA clouds depend on
      assumptions about their optical depth and the fraction of
      \ion{H}{i} emission that originates behind the cloud. After
      considering the most probable values for these parameters, Kavars
      et al. (\cite{Kav05}) found that the spin temperatures of the 70
      largest HISA complexes in their catalog range from 6 to
      $41~\mathrm{K}$.

      Density and mass estimates of the clouds depend on the same
      assumptions, made for calculating the spin temperatures, but also
      on the uncertainties in the distance estimates and in the filling
      factor of the cold gas in the clouds. For their 70 HISA complexes,
      Kavars et al. (\cite{Kav05}) find that the total hydrogen density
      may range from 42 to $550~\mathrm{cm}^{-3}$, and the \ion{H}{i}
      masses are about $3-4\,400~M_{\sun}$. Tighter constraints on the
      physical properties are possible for clouds that are seen to move
      from absorption into emission. Such an analysis for the HISA
      feature at $l=115\fdg5$, $b=7\degr$,
      $V_\mathrm{LSR}=-13~\mathrm{km\,s}^{-1}$ is done by Kerton
      (\cite{Ker05}), for example. As a result, he reported the spin
      temperature $30.5 < T_\mathrm{s} < 45~\mathrm{K}$ and the optical
      depth $0.97 < \tau < 3.7$. For this cold \ion{H}{i} feature he
      also estimated the column density $N = (1.0-6.9) \times
      10^{20}~\mathrm{cm}^{-2}$, the number density
      $51-260~\mathrm{cm}^{-3}$, and the total mass of the object $M =
      5-48~M_{\sun}$.

      To distinguish the HISA features from the gaps in the \ion{H}{i}
      emission profiles, which are caused by other factors than
      absorption, the detected HISA features were verified in early
      studies by molecular or dust tracers at the same velocity and
      position (Knapp \cite{Kna74}; Baker \& Burton \cite{Bak79}). This
      was justified by the common presumption that HISA gas is too cold
      to exist without some form of molecular cooling and shielding from
      the interstellar radiation field (Gibson \cite{Gib11}), and it led
      to the conclusion that the HISA features are caused by the
      residual amounts of very cold \ion{H}{i} in molecular clouds
      (Burton et al. \cite{Bur78}; Baker \& Burton \cite{Bak79}; Burton
      \& Liszt \cite{Bur81}; Liszt et al. \cite{Lis81}). Later the cold
      \ion{H}{i} clouds were found in which the \ion{H}{i} distribution
      only in part coincides with that of CO cloud, while the fragments
      of cool \ion{H}{i} are found well beyond the limit of the CO
      detection (Hasegawa et al. \cite{Has83}; Gibson et al.
      \cite{Gib00}; Kavars et al. \cite{Kav05}), or corresponding CO
      emission is missing at all (Peters \& Bash \cite{Pet83,Pet87}).
      Therefore, when the automated methods of feature identification
      and extraction for new large-scale Galactic plane surveys were
      developed, the requirement of verification by molecular or dust
      tracers was abandoned (Gibson et al. \cite{Gi05b}; Kavars et al.
      \cite{Kav05}).

      The results for the Galactic distribution of HISA and their
      relation to molecular gas are summarized by Gibson (\cite{Gib11}).
      He points out that weak self-absorption is found essentially in
      all directions where the emission background is bright enough.
      However, stronger HISA is clumped into complexes along spiral arms
      and tangent points. HISA also has a varying degree of
      correspondence with the CO emission. Most of the inner-Galaxy HISA
      has matching CO, but most outer-Galaxy HISA does not. In the SGPS
      only about 60\% of the identified HISA clouds have a
      $^{12}\mathrm{CO}$ brightness temperature of at least
      $0.5~\mathrm{K}$ (Kavars et al. \cite{Kav05}), but some HISA
      lacking CO show far-infrared dust emission. Many HISA features
      have a filamentary appearance (Taylor et al. \cite{Tay03}), but
      there are also examples of cometary HISA clouds (see Fig.~2 in
      Gibson \cite{Gib11}). Hosokawa \& Inutsuka (\cite{Hos07}) have
      found shell-like HISA features around the giant \ion{H}{ii}
      regions W4 and W5. Even a more spectacular feature is the cold,
      dark arc of Knee \& Brunt (\cite{Kne01}).

   \section{Infrared and submillimeter structures}

      With its unprecedented sensitivity and broad spectral coverage in
      the submm-to-mm range, the full-sky survey performed by the Planck
      satellite is providing an inventory of the cold clumps (CC) of the
      interstellar matter in the Galaxy (Planck Col. \cite{Pl11b}).
      However, the detection method used to extract sources from the
      Planck data is based on the color signature of the objects. This
      results in the discovery of more extended cold components with
      more complex morphology than the sources found with methods that
      identify structures on the basis of surface brightness (Planck
      Col. \cite{Pl11a}). Because the separation of NHIE objects was at
      least partly based on the surface brightness distribution, it is
      clear that comparing their properties with those of the Planck CC
      may be somewhat problematic. Moreover, the Planck CC are most
      likely a heterogeneous ensemble of objects, in which only the
      smallest nearby sources are probably the cold cores. Most of the
      others trace cold dust in larger irregular structures up to the
      mass of giant molecular complexes, and a small fraction of the
      sources in the Galactic plane may be superpositions of nearby and
      distant sources (Planck Col. \cite{Pl11b}). Therefore, when trying
      to compare the properties of the Planck CC with the parameters of
      NHIE clouds, we focus our attention on the nearby subsample of CC
      (Figs. 17 and 18 of the Planck Col. \cite{Pl11b}) or on the
      smallest regions of higher resolution observations (Table~3 of the
      Planck Col. \cite{Pl11a}).

      \begin{table*}
      \caption{Properties of cold clouds}
      \label{Tab02}
      \centering
      \begin{tabular}{l c c c c}
      \hline\hline
      & NHIE & HISA & Planck CC & IRDC \\
      \hline
      Complex size (pc) & $\sim50$ & $10-110$ & $<20$ & $<30$ \\
      Cloud size (pc) & $0.3-4.1$ & $0.029-3.6$ & $0.2-2.4$ & $0.4-1.0$ \\
      Temperature (K) & $12-47$ & $6-45$ & $7-21$ & $8-28$ \\
      Column density ($\times 10^{20}~\mathrm{cm}^{-2}$) & $2.5$ & $1.0-6.9$ & $8-160$ & $50-200$ \\
      Number density ($\mathrm{cm}^{-3}$) & $150-320$ & $42-550$ & $5 \times 10^3 - 10^5$ & $5 \times 10^5-3 \times 10^6$ \\
      Mass ($M_{\sun}$) & $0.235-1.07$ & $5-48$ & $0.4-1\,800$ & $23-300$ \\
      \hline
      \end{tabular}
      \end{table*}

      Whole-sky catalogs exist both for NHIE and the Planck CC, but the
      data do not have very good resolution, and the catalogs may
      contain rather heterogeneous ensembles of objects. For \ion{H}{i}
      more detailed absorption observations exist (HISA). The cold
      structures, observed in absorption at shorter wavelengths are
      infrared dark clouds (IRDC). These clouds were initially
      discovered by the Infrared Space Observatory (ISO, Perault et al.
      \cite{Per96}) and the Midcourse Space Experiment (MSX, Carey et
      al. \cite{Car98}; Egan et al. \cite{Ega98}) as dark structures
      against the bright mid-infrared background of the Galaxy. Later
      extensive catalogs of IRDC have been compiled by Simon et al.
      (\cite{Si06a}) and Peretto \& Fuller (\cite{Per09}). These clouds
      may be closely related to the Planck CC because fewer than 8\% of
      the Planck clumps inside the region studied by Simon et al.
      (\cite{Si06a}) are not directly associated with IRDC, and the
      Planck observations are sensitive to lower dust column densities
      than those of MSX (Planck Col. \cite{Pl11b}). From the comparison
      of the physical parameters of the Planck CC and IRDC, the Planck
      Col. (\cite{Pl11a}) has proposed that in general the Planck CC
      population may be representative of a slightly earlier stage of
      the evolution of IRDC cold dense cores.

      From the Planck Col. (\cite{Pl11a,Pl11b}) we may conclude that the
      Planck CC are observed in the distance range of
      $0.14-7.0~\mathrm{kpc}$, their linear dimensions are about
      $0.2-2.4~\mathrm{pc}$, and the dimensions of the cloud complexes
      extend up to $20~\mathrm{pc}$. The estimated color temperatures of
      the CC are mostly in the range of about $7-15~\mathrm{K}$, but
      some substructures are found to be warmer with $19-21~\mathrm{K}$.
      For such clumps, the presence of bright compact sources within the
      Planck-detected structures has been revealed (Juvela et al.,
      \cite{Juv10}). These sources are probably very young stellar
      objects, still embedded in their cold surrounding cloud. The
      column density, number density, and the total mass estimates for
      the Planck CC are $(0.8-16) \times 10^{21}~\mathrm{cm}^{-2}$, $5
      \times 10^3 - 10^5~\mathrm{cm}^{-3}$, and $0.4-1\,800~M_{\sun}$,
      respectively.

      IRDC are preferably high column-density objects at the distances
      up to $8.0~\mathrm{kpc}$ (Kainulainen et al. \cite{Kai11}). The
      typical linear size of an IRDC is about $5~\mathrm{pc}$ with some
      larger ones extending up to $30~\mathrm{pc}$ (Simon et al.
      \cite{Si06b}). IRDC usually contain smaller cores, defined as
      localized regions of higher extinction than the cloud's average
      (Simon et al. \cite{Si06a}). These cold, compact cores have
      typical sizes of about $0.5~\mathrm{pc}$ (Rathborne et al.
      \cite{Rat06}, Wilcock et al. \cite{Wil11}). Using the kinematic
      distances, Simon et al. (\cite{Si06b}) have estimated that IRDC
      have typical peak column densities of $\sim
      10^{22}~\mathrm{cm}^{-2}$, volume-averaged $\mathrm{H}_2$
      densities of $\sim 5 \times 10^3~\mathrm{cm}^{-3}$, and local
      thermodynamic equilibrium masses of $\sim 5\,000~M_{\sun}$. Many
      of the IRDC are completely opaque at wavelengths $7-100~\mu
      \mathrm{m}$. This lack of emission constrains the dust temperature
      to $<25~\mathrm{K}$. The median values of these parameters for
      cores are $\log[N(\mathrm{H}_2)(\mathrm{cm}^{-2})]=22.01 \pm
      0.29$, $\log[n(\mathrm{H}_2)(\mathrm{cm}^{-3})]=6.06 \pm 0.39$,
      and $\log[M(M_{\sun})]=1.92 \pm 0.55$ (Rathborne et al.
      \cite{Rat10}). The temperatures of the cores range from
      $8-11~\mathrm{K}$ at the center to $18-28~\mathrm{K}$ at the
      surface (Wilcock et al. \cite{Wil11}). The local temperature
      minima are strongly correlated with column density peaks, which in
      a few cases reach $10^{23}~\mathrm{cm}^{-2}$ (Peretto et al.
      \cite{Per10}). Many of the cores in IRDC are associated with
      bright $24~\mu \mathrm{m}$ emission sources, which suggests that
      they contain one or more embedded protostars. These active cores
      typically have warmer dust temperatures than the more quiescent,
      perhaps ``pre-protostellar'', cores (Rathborne et al. \cite{Rat10}).

      The distribution of the Planck CC is mostly concentrated in the
      Galactic plane, but some detections are also observed at high
      Galactic latitudes. The population is closely associated with
      Galactic structures, especially the molecular component: more than
      95\% of the clumps are associated with CO structures and about
      75\% are associated with an extinction greater than 1.
      Superimposed on the large-scale spiral structure of the Galaxy is
      a distribution of features known as shells, loops, etc. CC are
      primarily distributed on such structures (Planck Col.
      \cite{Pl11b}). The clumps are found to be significantly elongated
      and embedded in filamentary or cometary large-scale structures
      (Planck Col. \cite{Pl11a}). IRDC have been searched so far mostly
      in the inner Galaxy at $|l| < 60\degr$ and $|b| < 1\degr$. It has
      been found that they may represent the densest clumps within giant
      molecular clouds (Simon et al. \cite{Si06b}), and their
      distribution in the Galaxy may follow the spiral arms (Jackson et
      al. \cite{Jac08}).

   \section{Discussion}

      In this section we would like to compare the properties of the
      cold structures in the ISM, described in the previous sections,
      but this is not a straightforward task. Various methods are used
      to derive physical parameters from different kinds of
      observations, also resulting in a slightly different meaning of
      the obtained results. The observations used do not have the same
      angular resolution, and the observed objects are located from our
      local neighborhood to the outskirts of the Galaxy. Therefore, it
      is easy to confuse the small cores in the nearby objects with the
      complexes of such cores in more distant clouds or from the
      observations with poorer resolution. The mass and density
      estimates for NHIE and HISA are mostly from \ion{H}{i}, whereas
      these estimates for the Planck CC and IRDC are mainly from the
      dust and molecular data, etc. Nevertheless, we try to concentrate
      on the coolest and densest structures in the objects discussed,
      and to compare their temperatures, sizes, densities, and masses in
      the hope of revealing at least some general trends. A short
      compilation of such data from the previous sections of this paper
      is given in Table~\ref{Tab02}.

      If we compare NHIE and HISA, we may conclude that most of the
      properties of these two types of \ion{H}{i} clouds are at least
      very similar. The only obvious difference is in their sky
      distribution. When HISA features are observed only near the
      Galactic plane, our NHIE clouds do not even demonstrate a
      remarkable concentration on this plane. It may seem that the
      reason for the discrepancy is that we have looked for NHIE clouds
      in the all-sky \ion{H}{i} survey, but HISA features are searched
      for in the Galactic plane surveys, which do not extend to high
      latitudes. However, even in the narrow strips of these plane
      surveys, a strong concentration of HISA in the Galactic plane is
      clearly visible (e.g. Fig.~1 of Gibson et al. \cite{Gi05a}).

      Actually, the discrepancy in the sky distributions of HISA and
      NHIE clouds is a reflection of the differences in the observing
      conditions of these features. For self-absorption the background
      brightness temperature needs to be higher than the spin
      temperature of the absorbing gas. As most of the gas is
      concentrated in the Galactic plane, here we have plenty of both
      the foreground clouds and background \ion{H}{i} emission.
      Therefore, near the Galactic plane may we expect to find a lot of
      HISA features. At higher latitudes the general gas density
      decreases quickly, and even if there are cold \ion{H}{i} clouds,
      they do not have enough background emission to be observed in
      absorption. If at all, such cold clouds could be found there as
      NHIE features. An additional factor that reduces the concentration
      of our NHIE clouds to the Galactic plane, may be that the results
      of the Gaussian decomposition are more questionable in regions of
      more complicated \ion{H}{i} profiles, where different features are
      heavily blended by each other. Therefore, very narrow Gaussian
      components are harder to detect at lower latitudes.

      This explanation of the differences in the sky distribution of
      NHIE and HISA features seems to some extent also supported by the
      comparison of our Fig.~\ref{Fig04} with Fig.~4 of Gibson
      (\cite{Gib11}). From the Gibson's figure we can see that the
      strongest HISA is observed in the I and IV quadrants of the
      Galaxy. These quadrants correspond to the inner Galaxy with high
      gas densities and double-valued distance-velocity relation. This
      means that with the high probability for any \ion{H}{i} cloud
      closer to us than the subcentral point of the sightline, there is
      enough background gas behind the subcentral point with the same
      velocity as that of the foreground cloud, so that the foreground
      cloud is seen in absorption against the emission of the background
      gas. In the outer Galaxy, the gas densities are generally lower,
      and the distance-velocity relation is single valued. In these
      regions, it is much harder to find suitable background sources for
      cold clouds to be observed in absorption, and therefore they are
      relatively rare in quadrants II and III. Here a possible source of
      background emission is discussed by Gibson et al. (\cite{Gi05a}).

      If the cold \ion{H}{i} clouds exist in the regions of the sky
      where they are hardly observable in absorption, they may be
      observable in emission as NHIE. From our Fig.~\ref{Fig04} we see
      that most of the NHIE are probably rather local, since they have
      low line-of-sight velocities. According to Fig.~3 of Paper~I or
      Fig.~\ref{Fig02} here, these clouds are also located at relatively
      high Galactic latitudes, where the background emission is weak.
      The sky distribution of the small NHIE clouds with
      $|V_\mathrm{LSR}| > 15~\mathrm{km\,s}^{-1}$ and $N < 7$ is more
      strongly concentrated in the Galactic plane, and most of them have
      $|b| < 10\degr$. However, as can be seen from Fig.~\ref{Fig04},
      these objects are mostly observed in Galactic quadrants II and
      III, where the cold clouds are less likely observed in absorption.
      Therefore, the sky distributions of the HISA and NHIE are largely
      complementary to each other as if they have been derived from the
      same space distribution of objects by mutually exclusive observing
      conditions.

      We do not see good agreement in the mass estimates of NHIE and
      HISA either. However, the estimate for NHIE is based on the
      properties of only one cloud (Verschuur's cloud A), which is
      clearly a small, nearby subcondensation in a considerably larger
      NHIE structure (the ring, discussed in Paper~I). Most of the HISA
      mass estimates discussed in the present paper are for large HISA
      complexes, but the value given in Table~\ref{Tab02} is also
      derived for only one feature, whose mass may have been estimated
      more reliably than the masses of other clouds, because this cloud
      is seen in transition from absorption to emission. At the same
      time, the dimensions ($35 \times 1.7~\mathrm{pc}$; Kerton
      \cite{Ker05}) of this HISA at the distance of about
      $1~\mathrm{kpc}$ are comparable to the full size of the observable
      part of our ring. If we suppose that this feature may also contain
      approximately the same number of subcondensations as the ring, and
      we divide its mass estimate by the number of assumed
      subcondensations, the agreement is much better. Moreover, if we
      compare the mass of Verschuur's cloud A, given in
      Table~\ref{Tab02}, for example with the $p = f_n = 1$ estimate for
      the Perseus HISA globule $M_\mathrm{HISA} = 0.53-0.80~M_{\sun}$
      (Gibson et al. \cite{Gi05b}), the agreement is very good.

      Finally, some cases exist where one part of the cold \ion{H}{i}
      cloud is seen in self-absorption and the other part in emission.
      An example of this situation is the Riegel \& Crutcher cold cloud,
      first reported as HISA by Heeschen (\cite{Hee55}) and afterwards
      studied by Riegel \& Jennings (\cite{Rie69}), Riegel \& Crutcher
      (\cite{Rie72}), and others. Montgomery et al. (\cite{Mon95})
      reported that at longitude $l = 9\fdg87$ this cloud is detected in
      self-absorption between $b = -4\fdg2$ and $+8\degr$, but outside
      this latitude range, the \ion{H}{i} is observed in emission at a
      similar velocity. In the corresponding region of the sky, we have
      also detected the NHIE cloud fragments with the similar velocity.
      The HISA/NHIE cloud, studied by Kerton (\cite{Ker05}), has too
      small an angular extent to be detected by our algorithm, but
      nearby we have found an NHIE cloud with rather similar parameters
      $l = 119\fdg7$, $b = 6\fdg4$, and $V_\mathrm{LSR} =
      -10.3~\mathrm{km\,s}^{-1}$. Recently Moss et al. (\cite{Mos12})
      have found a local Galactic supershell GSH~006-15+7 in which the
      transition from \ion{H}{i} emission to self-absorption is
      observed. Fragments of this supershell seem to also be visible on
      our NHIE map (Fig.~5 of Paper~I or Fig.~\ref{Fig02} here). All
      this leads us to the conclusion that most likely HISA and NHIE
      features both represent the same physical class of cold \ion{H}{i}
      clouds in different observing conditions.

      When we compare the sky distributions of the Planck CC and IRDC
      with those of NHIE and HISA, we must take similar considerations
      into account as explained when comparing NHIE and HISA. The
      catalog of NHIE clouds is based on the LAB data with the gridpoint
      separation of about $0\fdg5$ (Kalberla et al., \cite{Kal05}). Only
      the nearest clouds have large enough apparent sizes to be detected
      in at least seven gridpoints, as required for the catalog, and
      such clouds demonstrate only weak concentration in the Galactic
      plane. The angular resolution of the Planck observations is about
      $4.5\arcmin$ (Planck Col. \cite{Pl11b}), the proportion of the
      distant, apparently smaller clumps in the detected sample is
      higher, and they demonstrate clearer concentration to the Galactic
      plane, but there are also objects at high Galactic latitudes. The
      surveys of HISA mostly have the resolution of $1\arcmin$ (Taylor
      et al. \cite{Tay03}, Stil et al. \cite{Sti06}), and these
      observations also require the presence of bright background
      emission. Therefore, the observations were only made near the
      Galactic plane and the detections demonstrate strong concentration
      to it. The same is also true for IRDC, but for them the resolution
      of the data in sky coordinates is even better ($20\arcsec$ for
      MSX, up to $5\arcsec$ for the Herschel Infrared Galactic Plane
      Survey, and $2\arcsec$ for Spitzer satellite data). Nevertheless,
      all four classes of objects may be related to shock fronts in the
      ISM, in the spiral structure, or in shell-like structures. Also
      the shapes of these objects share similarities: the filamentary or
      cometary structures are often found.

      By other parameters, the Planck CC and IRDC are most likely
      different from NHIE and HISA. They seem to be even colder than
      most of NHIE and HISA, their linear sizes are slightly smaller,
      and densities considerably higher than those of \ion{H}{i}
      features. And, of course, when only about half of the HISA
      features seem to be related to the $^{12}\mathrm{CO}$ emission,
      practically all of the Planck CC and IRDC are dominated by the
      molecular gas.

      Gibson \& Taylor (\cite{Gib98}) pointed out that the complex forms
      of HISA exhibit morphological aspects of both HI emission wisps
      and molecular cloud clumps. We have already mentioned that many
      HISA features also appear to be associated with CO emission,
      though the wide range of CO brightness to HISA opacity precludes a
      simple relationship between the two. Gibson \& Taylor
      (\cite{Gib98}) concluded that a possible explanation for this is
      that we may be seeing the actual phase transition from atomic to
      molecular gas brought about by the shock environment. This means
      that HISA features may be an intermediate evolutionary phase
      between the diffuse CNM and much denser molecular clumps. This
      assumption was later elaborated by Kavars et al. (\cite{Kav05})
      and others. From the data presented in the present paper, it seems
      natural to suppose that NHIE clouds also represent the same
      intermediate phase between diffuse CNM and molecular clumps, later
      represented by the Planck CC and IRDC features.

      The description of the discussed objects as different evolutionary
      stages in the conversion of diffuse CNM to molecular clumps, in
      which new stars may be born, seems to also be in general agreement
      with some theoretical calculations. We have pointed out several
      times that the structures described in this paper seem to have a
      certain relation to shock fronts in the ISM. Hosokawa \& Inutsuka
      (\cite{Hos07}) have studied the role of an expanding \ion{H}{ii}
      region in the ambient neutral medium. They have found that a shock
      front emerges and sweeps up the ambient CNM. The swept-up shell
      becomes cold ($T \sim 30~\mathrm{K}$) and dense ($n \sim
      10^3~\mathrm{cm}^{-3}$), and $\mathrm{H}_2$ molecules form in the
      shell without CO molecules. This is just the intermediate phase
      between the neutral medium and molecular clouds, something like
      HISA or NHIE clouds in which $\mathrm{H}_2$ may already exist, but
      it is hardly detectable since CO is still largely missing. Later
      the shell will fragment into small clouds as a result of
      gravitational instability. If each fragment contracts into a dense
      core and increases the column density, CO molecules may form.

      Molinari et al. (\cite{Mol10}) have outlined a scenario where
      diffuse clouds first collapse into filaments, which later fragment
      to cores. They point out that recent MHD numerical simulations
      (Banerjee et al. \cite{Ban09}) of the formation and subsequent
      fragmentation of filaments in the post-shock regions of large
      \ion{H}{i} converging flows or in the context of helical magnetic
      fields (Fiege \& Pudritz \cite{Fie00}) are in good agreement with
      the first results from the Hi-GAL survey on the core-hosting
      filaments, as well as with the mass regime of the cores being
      formed.

      From the data presented in this paper, only the comparison of the
      masses of NHIE and HISA clouds with those of the Planck CC and
      IRDC cores may be somewhat disturbing. The estimates for the
      Planck CC and IRDC seem to be considerably higher than those for
      NHIE and HISA. However, we must consider at least two
      circumstances. First of all, the estimates for NHIE and HISA are
      mostly based only on \ion{H}{i}, and so the possible contribution
      of the molecular gas remains largely unknown. Moreover, the cores
      of IRDC (and recalling their correlation with the Planck CC, also
      the latter) are considered as the places of high-mass star
      formation (Rathborne et al. \cite{Rat06}). The earliest phase of
      isolated low-mass star formation occurs within Bok globules
      (Rathborne et al. \cite{Rat10}). Viewed against background stars,
      Bok globules are identified as isolated, well-defined patches of
      optical obscuration (Bok \& Reilly \cite{Bok47}). The cores of Bok
      globules are typically small ($\sim 0.05~\mathrm{pc}$) and dense
      ($10^5-10^6~\mathrm{cm}^{-3}$), with low temperatures ($\sim
      10~\mathrm{K}$) and low masses ($0.5-5~M_{\sun}$; e.g., Myers \&
      Benson \cite{Mye83}; Ward-Thompson et al. \cite{War94}). Many of
      these low-mass star-forming regions are nearby, as most likely are
      also our NHIE clouds, and their masses are of the same order of
      magnitude as the mass estimate of Verschuur's cloud A.

   \section{Conclusions}

      In this paper, we have argued that the Gaussian decomposition
      enables us to separate narrow line components from the LAB
      \ion{H}{i} 21-cm line database and to compile a list of candidates
      of the NHIE clouds. To exclude artificial clouds, caused by RFI,
      non-uniqueness of the Gaussian decomposition in heavily blended
      cases, and the observational noise, we used the comparison of the
      LAB results with those obtained from the GASS in the southern sky.
      The obtained lists of NHIE cloud candidates for the LAB and GASS
      are available on request from the author (urmas@aai.ee). Then we
      reviewed the sizes, gas temperatures, column and number densities,
      and masses for NHIE and HISA clouds and for the Planck CC and
      IRDC. We also discussed the distribution of these clouds in the
      Galaxy, their basic shapes, and composition. From this discussion
      we draw the following conclusions
      \begin{itemize}
         \item The LAB Survey enables us to compile the low-resolution
            all-sky catalog of NHIE cloud candidates.
         \item NHIE objects share the physical properties of HISA clouds
            and may therefore be the same type of clouds as HISA, but in
            different observing conditions. In some respects, these
            clouds resemble the diffuse CNM, but NHIE and HISA are
            denser and colder.
         \item The Planck CC and IRDC are even colder and denser, and
            they contain more molecular gas than NHIE and HISA clouds,
            but by their distribution in the Galaxy and shapes they
            still resemble NHIE and HISA features.
         \item As proposed by Gibson \& Taylor (\cite{Gib98}) and Kavars
            et al. (\cite{Kav05}) for HISA features, NHIE clouds may
            also be an intermediate phase between the diffuse cold
            neutral medium and molecular clumps, represented by the
            Planck CC and IRDC or Bok globules for the nearby, less
            massive clumps.
      \end{itemize}

      To obtain more detailed whole-sky information on the NHIE clouds,
      new large-scale \ion{H}{i} 21-cm line surveys with better angular
      and velocity resolution than that of the LAB are needed.
      Unfortunately, the next generation whole-sky surveys, such as the
      GASS and EBHIS, offer better angular resolution than the LAB, but
      the velocity resolution is essentially the same. Both resolutions
      are considerably better in GALFA-\ion{H}{i} (the Galactic Arecibo
      L-band feed array \ion{H}{i}; Peek et al. \cite{Pe11a}) survey,
      but it can cover only about a $39\degr$ wide strip across the sky.

   \begin{acknowledgements}
      The author would like to thank M.~Walmsley for initiating the
      discussion on the relation between NHIE and HISA clouds, E.~Saar
      for asking for the comparison of the properties of NHIE features
      and the Planck cold clumps, and P. M. W. Kalberla for preparing
      the GASS for the decomposition and for discussing the comparison
      of the LAB and GASS results. I am also grateful to M.~Einasto,
      J.~Pelt, and P.~Hein\"am\"aki for encouraging me to write this
      paper and W. B. Burton for serving as a referee of the paper and
      suggesting valuable improvements. The project was supported by the
      Estonian Science Foundation grant no. 7\,765.
   \end{acknowledgements}

\end{document}